\documentclass[journal,12pt]{IEEEtran}

\usepackage{ifpdf}
\usepackage{flushend}
\usepackage{cite}
\usepackage{textgreek}
\usepackage[pdftex]{graphicx}
\ifCLASSINFOpdf
\else
\fi

\usepackage[cmex10]{amsmath}
\usepackage{amsfonts}
\usepackage{amsthm}

\usepackage{algorithm}
\usepackage{algorithmic}
\usepackage{mdwmath}
\usepackage{mdwtab}
\usepackage{eqparbox}
\usepackage[tight,footnotesize]{subfigure}
\usepackage{url}
\usepackage{tabularx}
\usepackage{booktabs}
\usepackage{multirow}

\usepackage{color}

\usepackage{svg}

\usepackage{tabularx}
    \newcolumntype{L}{>{\raggedright\arraybackslash}X}
\usepackage[pdfpagelabels,hypertexnames=false,breaklinks=true,bookmarksopen=true,bookmarksopenlevel=2]{hyperref}

\theoremstyle{plain}

\begin{document}

\markboth{Accepted for publication in the IEEE COMMUNICATIONS STANDARDS MAGAZINE}{Left}

\title{A New Agent-Based Intelligent Network Architecture}
%{Agent Based Unifed Grand Architectural Model for SDN, NFV, ETSI GANA, and ETSI MEC}
%{SDN Controller Decomposition and Multi-Agent Based Implementation}

\author{Sisay~Tadesse~Arzo, %~\IEEEmembership{Fellow,~OSA,}
        Domenico~Scotece, %~\IEEEmembership{Student Member,~IEEE,}
        Riccardo~Bassoli, %~\IEEEmembership{Senior Member,~IEEE,}
        Fabrizio~Granelli, %~\IEEEmembership{Senior Member,~IEEE,}
        Luca~Foschini %~\IEEEmembership{Senior Member,~IEEE,}
        and Frank~H.P.~Fitzek %~\IEEEmembership{Senior Member,~IEEE,}
\thanks{S. T. Arzo and F. Granelli are with the Department of Information Engineering and Computer Science (DISI), University of Trento, Trento, Italy.
(e-mail:  \{sisay.arzo,fabrizio.granelli\}@unitn.it).}%
\thanks{D. Scotece and L. Foschini are with Department of Information Engineering and Computer Science (DISI) at University of Bologna, Bologna, Italy. 
(e-mail: \{domenico.scotece,luca.foschini\}@unibo.it.}%
\thanks{R. Bassoli and F. H.P. Fitzek are with the Deutsche Telekom Chair of Communication Networks, Institute of Communication Technology, Faculty of Electrical and Computer Engineering, Technische Universität Dresden, Dresden, Germany.
\par F. H.P. Fitzek is also with Centre for Tactile Internet with Human-in-the-Loop (CeTI), Cluster of Excellence, Dresden, Germany.
(e-mail: \{riccardo.bassoli,frank.fitzek\}@tu-dresden.de).}%
%\thanks{Manuscript received ; revised .}
\thanks{{© 2022 IEEE. Personal use of this material is permitted. Permission from IEEE must be obtained for all other uses, in any current or future media, including reprinting/republishing this material for advertising or promotional purposes, creating new collective works, for resale or redistribution to servers or lists, or reuse of any copyrighted component of this work in other works.}}
}

\maketitle

\begin{abstract}
The advent of 5G and the design of its architecture has become possible because of the previous individual scientific works and standardization efforts on cloud computing and network softwarization. Software-defined Networking and Network Function Virtualization started separately to find their convolution into 5G network architecture. Then, the ongoing design of the future beyond 5G (B5G) and 6G network architecture cannot overlook the pivotal inputs of different independent standardization efforts about autonomic networking, service-based communication systems, and multi-access edge computing. This article provides the design and the characteristics of an agent-based, softwarized, and intelligent architecture, which coherently condenses and merges the independent proposed architectural works by different standardization working groups and bodies. This novel work is a helpful means for the design and standardization process of the futureB5G and 6G network architecture.

\end{abstract}

\section{Introduction}
The International Telecommunication Union (ITU) Telecommunication Standardization Sector (ITU-T) Focus Group on IMT-2020 concluded its pre-standardization activities in December 2016. The core reform that 5G introduced in communication was the idea of virtualization, or more specifically, network softwarization. This implied a tremendous paradigm shift from the previous store-and-forward to the current-future compute-and-forward. This has made computing as important as communication in future communication networks~\cite{CompBook}. 
However, the design of 5G architecture and core characteristics has leveraged the previous experience, obtained during the successful development of cloud computing and network virtualization instances such as software-defined networking (SDN) -- led by the no-profit consortium Open Network Foundation (ONF) -- and network function virtualization (NFV) -- started by the industry and standardized by the European Telecommunications Standards Institute (ETSI).

Recently, ITU has started writing a report about the future technology trends towards 2030 and beyond, which is going to be released in June 2022. This report will mainly provide the very general vision for future 6G communication networks. \textcolor{blue}{}In 2021, 6G research and design started in the EU, like the EU flagship Hexa-X (hexa-x.eu) and the German research hub 6G-life (6g-life.de). Especially, the Hexa-X project has been publishing the main guidelines for 6G characteristics, use cases, key performance indicators (KPIs), and architecture \cite{Hexa-X2021,Hexa-X2021arch}. However, the last few years of research and standardization efforts in in-network intelligence and network softwarization have been providing the mature input for the architectural aspects of the new generation, which is still an open challenge in \cite{Hexa-X2021,Hexa-X2021arch}. Thus, this article provides an architectural design and guidelines for future 5G, beyond 5G, and 6G networks by leveraging the various standardization works by ETSI and 3GPP, considering the current popular trends on microservice- and agent-based intelligent communication systems. In this vision, all softwarized network functions become atomic entities, playing as autonomous, intelligent, and collaborative agents\textcolor{blue}{}, in line with the current trends shown in \cite{Hexa-X2021,Hexa-X2021arch}. To the best of the authors' knowledge, this is the first architectural attempt to unify the four existing individual architectures, microservices-based SDN control plane, ETSI SDN-NFV Management and Orchestration (MANO), ETSI Generic Autonomic Networking Architecture (GANA), and 3rd Generation Partnership Project (3GPP)-ETSI Mobile Edge Computing (MEC), a unique and homogeneous framework for 5G, beyond 5G, and 6G.

\section{Motivation and Background}

%This section provides an overview of microservice-based SDN controller's architecture, and of the ETSI NFV, ETSI GANA, and 3GPP and ETSI MEC architecture and their characteristics. The section is intended to use to understand how the various standards concepts can be integrated in a unified architecture.

This section provides an overview of the main existing independent architectures. %adopting microservice-based communication systems, autonomic networking, and in-network intelligence.

\begin{figure*}
	    \centering%
	    {\includegraphics[width=1.8\columnwidth]{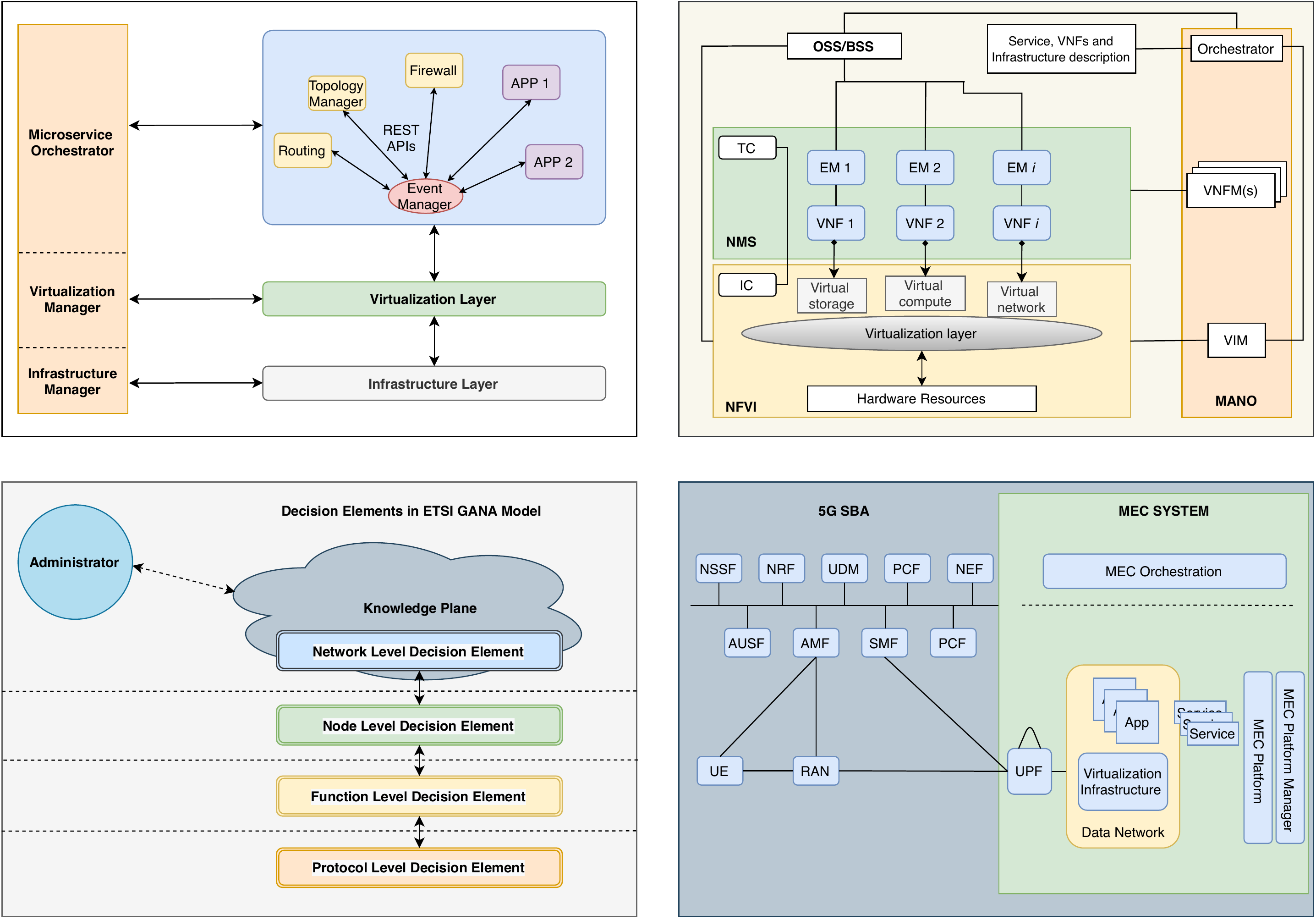}\hspace{0em}}%
	    \caption{\textcolor{blue}{}Architectures of microservices-based SDN Controller decomposition (top left), ETSI SDN-NFV MANO (top right), ETSI GANA (bottom left), and 3GPP-ETSI MEC (bottom right).}
	    \label{fig:SotAarchitecture}
\end{figure*} 

%This section presentes the motivation for the proposed approach a unified architecture that integrates intelligence in the network functions.  

%Dynamic response to hetroginous users with dyanmic demands. The widspeard application of machine learning (ML) for  

\subsection{Microservices-based SDN Controller}

\textcolor{blue}{}The centralized nature of the SDN control plane leads to several issues on latency, single point of failure, and overloading \cite{8187644}. These issues have been partially overcome by using the distributed SDN controller approach. However, the SDN controller is a monolithic system that results inefficient in replicating the SDN system. Moreover, monolithic SDN deployment does not allow dynamic management of SDN components and/or functionalities to (de-)activate according to the scenario. To alleviate this, the work in \cite{sdn-Decomposition} and \cite{microONOS} decomposed the SDN controller and implemented it as a set of microservices components that can run in a distributed fashion with flexibility in recomposition. In particular, the SDN controller is designed as a composition of logical sub-functions, i.e. microservices. They can share network service load and create a robust system against failures.  

%That represents a step towards the cloud continuum and is in line with the vision of the edge/cloud hybrid architecture \cite{cloud}.

\textcolor{blue}{}The work in  \cite{microONOS} presented \textmu ONOS architectural framework which is proposed by Open Network Operating System (ONOS) \cite{microONOS}. It is expected to be the next-generation SDN architecture. However, \textmu ONOS has been specialized mainly for cloud datacenters scenarios. It is limited to certain technologies and is not all 5G compliant. In particular, \textmu ONOS uses gRPC protocols family for microservices intercommunication which is not the standard declared in the 3GPP white-paper v15 \cite{3GPPv15}. 3GPP specifies REST APIs as the standard-de-facto for services intercommunication. In this regard, \textcolor{blue}{}we have implemented a microservices-based SDN controller deployment that complies with the 5G standard, see Figure 1 (top left) \cite{sdn-Decomposition}. \textcolor{blue}{}Our implementation is based on Ryu SDN Framework and a first release is available to the community at the link: \url{https://gitlab.com/dscotece/ryu_sdn_decomposition/}.

%\begin{figure}
%	    \centering%
%	    {\includegraphics[width=1.0\columnwidth]{Figures/microonos.png}\hspace{0em}}%
%	    \caption{\textmu ONOS Deployment Architecture}
%	    \label{fig:SDN}
%\end{figure}    
    
\subsection{ETSI SDN-NFV MANO}    

ETSI released it's ETSI SDN-NFV MANO architecture \cite{MANO}, to provide a unified architecture effectively combining SDN and NFV characteristics. Figure \ref{fig:SotAarchitecture} (top right) depicts its structure. This architecture consists of three main entities, called the Network Management System (NMS), the Network Function Virtualization Infrastructure (NFVI), and the Operation/Business Support Scheme (OSS/BSS). The first manages the virtual network, the second set the resources (hardware or software) that are used to run and to connect virtual network functions, and the third sets the applications used by service providers to provide network services.
Each one of these layers has an interface to the MANO entity, which hosts the virtual infrastructure managers (VIMs), that aim at controlling the NFVI resources. Next, the virtual network function manager (VNFM) configures and manages the life cycle of virtual network functions in its network domain. Finally, there is the orchestrator for the NFVI who manages the resources across different VIMs, and subsequently the life cycle of network services. Finally, the virtualization layer groups all the element management entities with their respective virtual network functions (VNFs).

\subsection{ETSI GANA}   

\subsubsection{Generic Autonomic Network Architecture}    
ETSI also unveiled a standard reference architecture as an implementation guide for GANA architectural framework for network automation \cite{ETSI}. The main goal of the GANA reference model is to prescribe the design and operational principles for Decision Elements (DEs) as the drivers for cognitive, self-managing, and self-adaptive network behaviors. ETSI is performing several GANA instantiations onto various target standardized reference network architectures. This is to enable autonomic algorithms to be integrated into the design and implementation of DEs. The integration also aimed at standardizing the autonomic network architectures at four levels as depicted in Figure 1 (bottom left). 

\textbf{Network Level Decision Element}: this level is designed to operate the outer closed control loops on the basis of network wide views or state as input to the DEs’ algorithms and logics for autonomic management.
\textbf{Node Level Decision Element}: this level is in charge of controlling the behavior of the Network Element (NE) as a whole, and also managing the orchestration and policing of the Function Level Decision Elements. GANA Function Level specifies the following four decision elements: security management, fault management, auto configuration and discovery, resilience, and survivability.
\textbf{Function Level Decision Element}: represents a group of protocols and mechanisms that abstract, as atomic units, networking or a management/control function. The GANA model defines the following six Function Level Decision Elements: routing management; forwarding management; Quality of Service management; mobility management; monitoring management; service and application management. 
\textbf{Protocol Level Decision Element}: this is the lowest level decision element in the system. This kind of element is protocols or other fundamental mechanisms that may exhibit intrinsic control-loops or decision element logic and associated DE, as is the case for some protocols such as Open Shortest Path First (OSPF). 
\textbf{GANA Knowledge Plane (GANA KP)}: enables advanced management and control intelligence at the Element Management (EM), Network Management (NM), and Operation and Support System (OSS).

\textcolor{blue}{}A close look at the GANA model reveals the close alignment with the multi-agent-based approach proposed in \cite{ArzoSis}, which defines the network functions as services that can be deployed in a virtualized/containerized environment.

\subsubsection{Multi-Agent System (MAS)}
is a sub-branch of distributed artificial intelligence, where it has multiple interacting intelligent agents performing a given task collaboratively and autonomously. MAS includes different attributes such as architecture, communication, coordination strategies, decision making, and learning abilities. The work in \cite{ArzoSis} identified MAS as a competing candidate for defining atomic and autonomous DEs in the GANA model. DEs could be network function units that could be used as a building block in any automated system. Network function atomization is the mechanism that defines the smallest possible network function units in a service-oriented system. MAS is comparable to microservices but with more autonomy and proactive capability \cite{ArzoSis}. %A comparative analysis is presented in \cite{ArzoSis}. 

\subsection{3GPP-ETSI MEC Architecture}

The basic idea of MEC is to provide capabilities closer to the end-users to overcome mobile difficulties. This promotes a new three-layer device-edge-cloud hierarchical architecture, which is recognized as very promising for several application domains \cite{ETSIMEC}. With the advent of 5G networks, the MEC is one of the key technologies for supporting ultra-reliable and low-latency communications.

The 5G system architecture specified by 3GPP is designed to fit a wide range of use cases including IoT networks management. In particular, the 3GPP defines the Service-Based Architecture (SBA) for the 5G core network, whereby the control plane functionality and common data repositories are delivered by a set of interconnected Network Functions (NFs), each with authorization to access each other’s services \cite{3GPP5G}. The SBA framework provides the necessary functionality to authenticate the consumer and to authorize its service requests, as well as flexible procedures to efficiently expose and consume services. Moreover, ETSI MEC defines an API framework aligned with the SBA framework to provide a set of functionalities and services. \textcolor{blue}{}The network functions and the services are registered in a Network Resource Function (NRF), while MEC applications are registered in the MEC platform. 5G Network Exposure Function (NEF) acts as a centralized point for service exposure. The Network Slice Selection Function (NSSF) is the function that selects the user slice and allocates the necessary Access Management Functions (AMF). The Unified Data Management (UDM) function generates the 3GPP AKA authentication credentials, while the User Plane Function (UPF) is the configurable data plane. The resulting integrated architecture described in the white paper \cite{3GPP5G} is presented in Figure~\ref{fig:SotAarchitecture} (bottom right).

\section{Proposed Architecture}
This section discusses the unified architecture, providing conceptual analysis. 
%and Standardization of the System}

\subsection{Proposed Unified Architecture}
%This section presents our novel fully unified architecture for automated network systems. 
\textcolor{blue}{}Our architecture combines four different standards namely: SDN, ETSI NFV, ETSI GANA, and ETSI MEC, for proposing a virtualized network intelligent agents instead of VNFs or microservice (in case of SDN decomposition). Figure 2 shows our unified architecture. %showing the overall structure. 
While the infrastructure and virtualization layers are directly derived from NFV architecture, the MANO layer is modified to have internal components as intelligent orchestration agents. %so that the internal components of the MANO are defined to be intelligent orchestration agents.
\textcolor{blue}{}The application layer of NFV is significantly modified and contains the four layers of the ETSI GANA model. Therefore, the ETSI GANA divides the application layer of NFV into four layers that we further propose the ETSI GANA decision units to be intelligent agents. Hence, the application layer becomes a composition of protocol-level agents, function-level agents, node-level agents, and network-level agents. The event distribution agents enable the event handling at each layer of the GANA model. Note that for event management several strategies are available including centralized, distributed, and hybrid \cite{hyperflow}. 

\textcolor{blue}{}The application layer depicts the decomposed controller, NMS, and MEC functions on the left and right sides, respectively. The functions are built as autonomous and atomic agents introducing in-network intelligence in the internal architecture of the agents. The recomposition of these agents would create the required controller, MEC system, and other NMS, depending on the requirements in a given environment. The recomposed system becomes a multi-agent-based intelligent system. Network functions in 5G, controller, and NMS and 5G/6G functions should be designed as autonomous agents incorporating appropriate AI/ML as an integral part of that function. The collection of such functions would create a fully autonomous network system. In other words, functions become intelligent agents where intelligence is introduced in the internal architecture of the agent, designing the agents using cognitive components such as AI/ML algorithms. 

\textcolor{blue}{}The interface between the infrastructure layer, virtualization layer, and the application layer is similar to NFV with little modification; however, the orchestration layer is based on intelligent agents. Therefore, we recommended interfaces to be open and adaptive based on the specific applications. This could be achieved by equipping agents with a Programmable Protocol Stack (PPS), which represents the implementation of a software-based environment of network protocols and layers \cite{PPS}. The interface between the agents is mainly through restful API. However, we suggested equipping the agent with a dynamic protocol stack.
  
\begin{figure*}
	\centering%
	{\includegraphics[width=1.8\columnwidth]{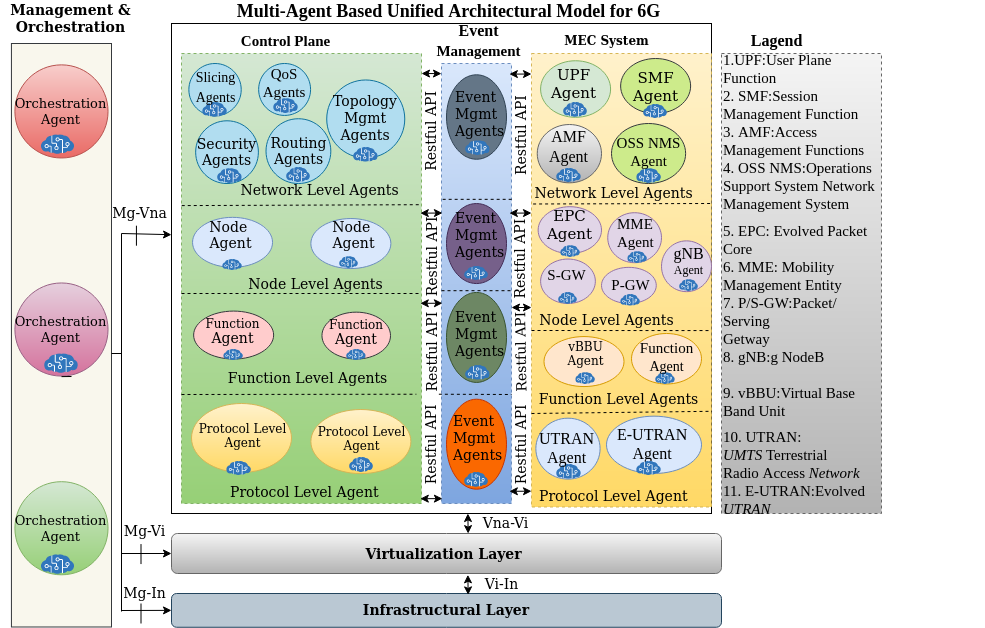}\hspace{0em}}%
	\textcolor{blue}{}\caption{Microservices and Multi-Agent Based Unified Architectural Model for ETSI GANA, ETSI MEC, ETSI NFV, and SDN}
	\label{fig:MicrONUS_NFV_GANA_MEC3}
\end{figure*}

\subsection{Agent Internal Architecture}
\begin{figure}
	    \centering%
	    {\includegraphics[width=0.7\columnwidth]{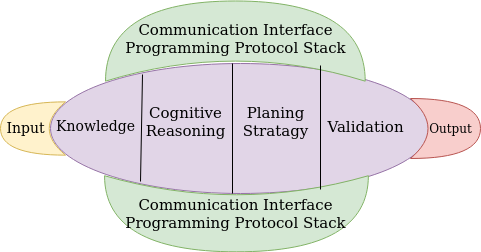}\hspace{0em}}%
	    \caption{Agent Internal Architecture}
	    \label{fig:SDN_Agent}
    \end{figure}
\textcolor{blue}{}We propose agent-designed principles and guidelines to enable intelligence and full autonomy in performing a given function incorporating cognitive components. Figure \ref{fig:SDN_Agent} depicts the general agent design and architectural guideline.  An agent is composed by an \textit{INPUT}, which is an incoming request; \textit{FACTS}, is the knowledge database of the agent; \textit{COGNITION (REASONING) UNIT} gives agents the reasoning capability; \textit{PLANNING STRATEGY} organizes the steps/procedures for the action to be taken to satisfy the requested service; \textit{VALIDATION} is the unit that verifies the action plan for consistency, and an \textit{OUTPUT/ACTION} is the final decision (result/action) to be taken by the agent. An agent also has an interface to communicate with other agents or microservices via the PPS. Future generation networks should be equipped with effective and efficient protocols. Agents in the application layer of the proposed architecture are various types that will use combinations of multiple protocols for communication for various kinds of applications including real-time applications. A PPS is a software-based stack that inputs information from the higher logical layers and configures various parameters at each layer, repeating the procedure at each network device. Parameters are updated dynamically according to network changes and service requirements.

%The concept of network function atomization is first discussed in \cite {4273111}, where the authors argued the idea for intelligent node organization. As per the original concept, every node in the system is equipped with an autonomic decision element which is the smallest object in an autonomic network. Besides, the idea in \cite {4575146} was a considering the composition of autonomic elements. This composition is intended to include an autonomous overlay network management structure and a self-organizing composition against autonomous overlay networks.

\subsection{Network Function Atomization}
\textcolor{blue}{}Network functions can be divided into small functions allowing maximum recomposition freedom. Here we redefine these functions as atomic units having full autonomic capability to perform a given function such as security, QoS monitoring, AMF, SMF, etc. Our architecture re-defines the internal components of a decomposed monolithic system to be intelligent agents, designing the agents using the cognitive element as the brain of the agents to make an autonomous decision on a given function. \textcolor{blue}{}The agents are fully intelligent, atomic, and autonomous decision units that can be flexibly recomposed to create a fully autonomic network system. Intelligent refers to the capability of the agent in performing its function, adapting to the dynamic demands. The level of intelligence and capability is heavily reliant on the design of a particular agent. For example, a network traffic classifier and predictor agents can be designed using ML model\cite{9685568} as a cognitive component of the agent and it can classify and predict given incoming service traffic in order to allocate network resources accordingly depending on the traffic class and proactive allocation and scheduling agents' decision.
%And a system that uses such agents \cite{9685568}, that are ideally able to perfectly classify and predict given incoming service traffic, it could be possible to allocate network resources accordingly depending on the traffic class and proactive allocation and scheduling agents' decision. 
This reduces resource and energy consumption while reducing service processing latency using proactive resource allocation and scheduling. Again we are assuming task execution agents can optimize their performance depending on the current traffic and system state. Depending on the implemented ML model as a cognitive component of the agent, the performance of the agent varies in terms of accuracy, latency, etc\cite{9685568}. Similarly, an autonomous SMF agent, designed using an appropriate ML, can dynamically create, update, and remove protocol data unit sessions while managing session context with the UPF agent. Agent atomicity refers to the smallest functionality, identity, and acting territory of a particular function/agent. Autonomy defines the agent's self-reliance in delivering the given task without external intervention. However, the agent can communicate with other agents when it requires information or if completing the task requires the support of other agents. In general, the recomposition of agents would create the overall autonomous system. The recomposition could be agent chaining creation using an orchestration agent. This by itself require intelligence that should be incorporated when designing an intelligent orchestration agent.

%
%\begin{figure}
%	    \centering%
%	    {\includegraphics[width=1.0\columnwidth]{Figures/Agnet_Template.png}\hspace{0em}}%
%	    \caption{Agent Instantation From a Template}
%	    \label{fig:Agent_Template}
%   \end{figure}
    
\subsection{Multi-Agent Based System Recomposition}
%It is possible to design the SDN controller as a single monolithic process, as a confederation of identical processes arranged to share the load or protecting each other from failures, or as a set of distinct functional and collaborative components. Moreover, any combination of these alternatives is possible. 
\textcolor{blue}{}As a guideline in designing the agent-based system, we show a decomposed SDN controller\cite{sdn-Decomposition}, designing the components as autonomic agents, and recomposing them to create the intelligent controller system. %This is a big advantage in the era of functions' \textit{containerization}, and \textit{cloudification}, where features of loose-coupling and distributed deployment are required. %\cite{sdn-container}. 
The main principle to retain in decomposing an SDN controller is that the network information and state should be synchronized and self-consistent. This allows an independent implementation and component reuse. %In \cite{sdn-Decomposition} the authors  showed SDN controller decomposition using microservices.
We showed in \cite{sdn-Decomposition} a possible implementation of decomposed SDN Controller using microservice, however, now we re-define these components with intelligent and atomic agents. %Here we replace and re-define these components with intelligent and atomic agents. 
These agents can be executed on arbitrary computing platforms on distributed and virtualized resources such as virtual machines/containers in edge/cloud environments using intelligent orchestration agents. The loosely-composed system can be viewed as a black box, defined by its externally observable behavior, emulating the original monolithic SDN controller with added intelligence. Agents' orchestration can be done using intelligent orchestrator agents. This creates an agent chain to intelligently perform what the legacy controller does. The recomposition can be done with flexible agent composition as per the system requirement in time and space. In particular, the controller could be deployed with maximum recomposition freedom using only the necessary functions/agents while performing the functions autonomously.

\section{Conclusion}
This paper presented a unified architecture combining SDN, NFV, MEC, and GANA architectural and conceptual models for future networks, such as beyond 5G, and 6G. \textcolor{blue}{}The main principles addressed by the architecture are the introduction of in-network intelligence using intelligent agents, decomposing monolithic network systems. Moreover, it also shows the organization of agents to produce an overall automated system. Due to the maximum recomposition freedom and in-network intelligence at the smallest service units of a network system, the architecture is expected to address future network demands in terms of autonomy, flexibility, heterogeneity, reliability, and latency. Since there are few agent based service design one of future challenges this architecture is developing such functions.

\section*{Acknowledgements}
This work is partially funded by NATO Science for Peace and Security Programme in the framework of the project SPS G5428 ”Dynamic Architecture based on UAVs Monitoring for Border Security and Safety”. This work is also partially funded by the German Research Foundation as part of Germany’s Excellence Strategy – EXC2050/1 – Project ID 390696704 – Cluster of Excellence “Centre for Tactile Internet with Human-in-the-Loop” (CeTI) of Technische Universität Dresden, and by the European Commission through the H2020 projects Hexa-X (Grant Agreement no. 101015956).

\ifCLASSOPTIONcaptionsoff
  \newpage
\fi

\bibliographystyle{IEEEtran}
\bibliography{References/refs}

\begin{IEEEbiographynophoto}{Sisay T. Arzo} is a postdoc at the Univ of New Mexico, USA. He received his PhD and MSc. in Telecom Eng from Uni of Trento. He received his BSc. from Hawassa Univ., Ethiopia. He has five years of industrial experience in Telecom industry. His research interest included network softwarization, IoT and Network Automation. 
\end{IEEEbiographynophoto}

\begin{IEEEbiographynophoto}{Domenico Scotece} is a postdoc researcher at the Univ of Bologna. He received his  Ph.D. and M.Sc. degrees in Computer Science Eng. from Univ of Bologna in 2020 and 2014,respectivly. His research interests include pervasive computing, middleware for fog and edge computing, IoT, and management of cloud computing systems.
\end{IEEEbiographynophoto}

\begin{IEEEbiographynophoto}{Riccardo Bassoli}
is  a  senior  researcher  with  the  Deutsche  Telekom  Chair  of  Communication  Networks  at  the  Faculty  of  Electrical and Computer Engineering, Technische Universität Dresden (Germany). He received his Ph.D. degree from 5G Innovation Centre (5GIC) at University of Surrey (UK), in 2016.
\end{IEEEbiographynophoto}

\begin{IEEEbiographynophoto}{Fabrizio Granelli}
Fabrizio Granelli is Full Professor at the Dept. of Information Eng. and Computer Science (DISI) of the Univ. of Trento,Italy. He was IEEE ComSoc Distinguished Lecturer for 2012-15 and 2021-22, Director for Online Content in 2016-17 and Director for Educational Services in 2018-19. He is author of more than 250 papers on networking.

\end{IEEEbiographynophoto}

\begin{IEEEbiographynophoto}{Luca Foschini} is Ass. Prof. of computer eng. His interests span from integrated management of distributed systems and services to edge computing, from management of cloud computing systems to Industry 4.0. He is a voting member of the IEEE ComSoc EMEA Board.
\end{IEEEbiographynophoto}

\begin{IEEEbiographynophoto}{Frank H.P. Fitzek}
received  the Dipl.-Ing. degree in electrical eng. from the Rheinisch-Westfälische Technische Hochschule, Aachen, Germany, in 1997 and the Dr.-Ing. degree in Electrical Engineering  from  the Technical University Berlin, Germany. He is  a  Professor  and  head  of the  Deutsche Telekom Chair of Communication Networks at Technische Universität Dresden.
\end{IEEEbiographynophoto}

%%%\begin{thebibliography}{1}
%%\end{thebibliography}

%\begin{IEEEbiography}{Riccardo Bassoli}
%\end{IEEEbiography}

%\begin{IEEEbiographynophoto}{XXX yyy}

%\begin{IEEEbiographynophoto}{XXX yyy}

\end{document}